\documentclass[%
reprint, 
 amsmath,amssymb,
 aps,
]{revtex4-2}

\usepackage[T1]{fontenc}
\usepackage{graphicx}
\usepackage{dcolumn}
\usepackage{bm}
\usepackage{mathtools}
\usepackage{caption}
\usepackage{subcaption}
\usepackage{xcolor}
\usepackage{comment}

\begin{document}

\title{Spatiotemporal Activity-Driven Networks}

\author{Zsófia Simon}
\email{zsofia.simon@aalto.fi}
\author{Jari Saramäki}
\affiliation{Department of Computer Science, Aalto University, Espoo 00076, Finland
}

\begin{abstract}
Temporal-network models have provided key insights into how time-varying connectivity shapes dynamical processes such as spreading. Among them, the activity-driven model is a widely used, analytically tractable benchmark. Yet many temporal networks, such as those of physical proximity, are also embedded in space, and spatial constraints are known to affect dynamics unfolding on the networks strongly. Despite this, there is a lack of similar simple and solvable models for spatiotemporal contact structures. Here, we introduce a spatial activity-driven model in which short-range contacts are more frequent. This model is analytically tractable and captures the joint effects of space and time. We show analytically and numerically that the model reproduces several characteristic features of social and contact networks, including strong and weak ties, clustering, and triangles having weights above the median.
These traits can be attributed to space acting as a form of memory.
Simulations of spreading dynamics on top of the model networks further illustrate the role of space, highlighting how localisation slows down spreading. 
Furthermore, the framework is well-suited for modelling social distancing in a principled way as an intervention measure aimed at reducing long-range links. We find that, unlike for non-spatial networks, even a small spatially targeted reduction in the total number of contacts can be very effective.
More broadly, by offering a tractable framework, the model enables systematic exploration of dynamical processes on spatiotemporal networks.

\end{abstract}

\maketitle
\section{Introduction}

In many complex systems, interactions between the constituent elements vary over time — consider, e.g., social interactions, transport, or networks of physical proximity. In such systems, dynamical processes including spreading, synchronisation, or information diffusion are strongly affected by their temporal structure~\cite{holme2012temporal,masuda2017temporal,TNT}. There are several reasons for this, from processes having to follow time-respecting paths to the effects of various types of temporal correlations and heterogeneity, such as burstiness~\cite{PhysRevLett.98.158702,Karsai2011,unicomb2021dynamics}. 

The activity-driven model is a foundational model for temporal networks~\cite{perra2012activity}. It was designed to capture the temporality and burstiness of real-world interactions as well as the heterogeneity in the number of contacts of individuals. This is achieved through activity potentials that vary in the population. The strength of this framework is that it is analytically tractable \cite{perra2012activity, pozzana2017epidemic, liu2014controlling}, which allows one to develop theory and intuition, as well as to test dynamical predictions, such as the results of epidemic spreading. However, the original model has not been designed to reproduce certain properties of social networks, such as heterogeneous link weights, heavy triangles, and high clustering. In practice, these characteristics have been reproduced in subsequent models by reinforcement mechanisms or memory (see, e.g.,~\cite{laurent2015}). 

In addition to being constrained by time, many real-world networks are also strongly embedded in space~\cite{BarthelemyReview,Hens2019} — consider, e.g., networks of mobility and physical proximity~\cite{balcan2009,brockmann2011,brockmann2013,stehle2011,Copenhagen,Schlosser20}. In the case of contact networks, the simultaneous constraints of time and space have been accounted for at a coarse-grained level with metapopulation models~\cite{balcan2009,Ajelli2010,doi:10.1126/science.aba9757,Arenas20,jia2020,SorianoPanos22} where agents flow between fully-mixed populations in a spatially structured network. At the micro level, small-scale empirical data sets that record proximity in time have proven very fruitful~\cite{stehle2011,Copenhagen,barrat2021effect}. 
However, theoretical work addressing the micro-level foundations of dynamics on spatiotemporal networks is still lacking. In particular, there is a lack of simple, analytically tractable models of spatiotemporal networks in the spirit of the activity-driven model.

To fill this gap, in this paper we introduce the spatial activity-driven model, where spatiality is brought about by adjusting the temporal activity-driven link-formation process to be spatially preferential. The proposed model aims to showcase how local, proximity-based link formation influences spatiotemporal network structure. We show analytically that space---acting as an inherent memory dimension---induces in aggregated networks the emergence of social network characteristics such as heterogeneous yet limited link weights, high clustering, and triangles with weights above the median~\cite{Onnela2007NJP} without an explicit reinforcement mechanism. We verify these results with simulations and demonstrate the model's applicability to studies of epidemic spreading.  
This spatiotemporal framework is highly flexible, as the spatial dimensions need not represent physical proximity but can instead serve as an abstract embedding space of the network nodes.

In the following sections, we describe the modelling framework and then demonstrate its analytical tractability with results that are in good agreement with simulation data. We proceed by showcasing aggregated characteristics that match well with those of social networks and also provide a basis for applications to spreading dynamics. We demonstrate the constraining effects of space through simulated  Susceptible-Infectious-Recovered (SIR) spreading on both the spatial and original activity-driven models. We conclude with a discussion, which includes an outlook on more general interpretations of the embedding space.

\section{Model}

\begin{figure*}[t]
    \centering
    \includegraphics[width=1\linewidth]{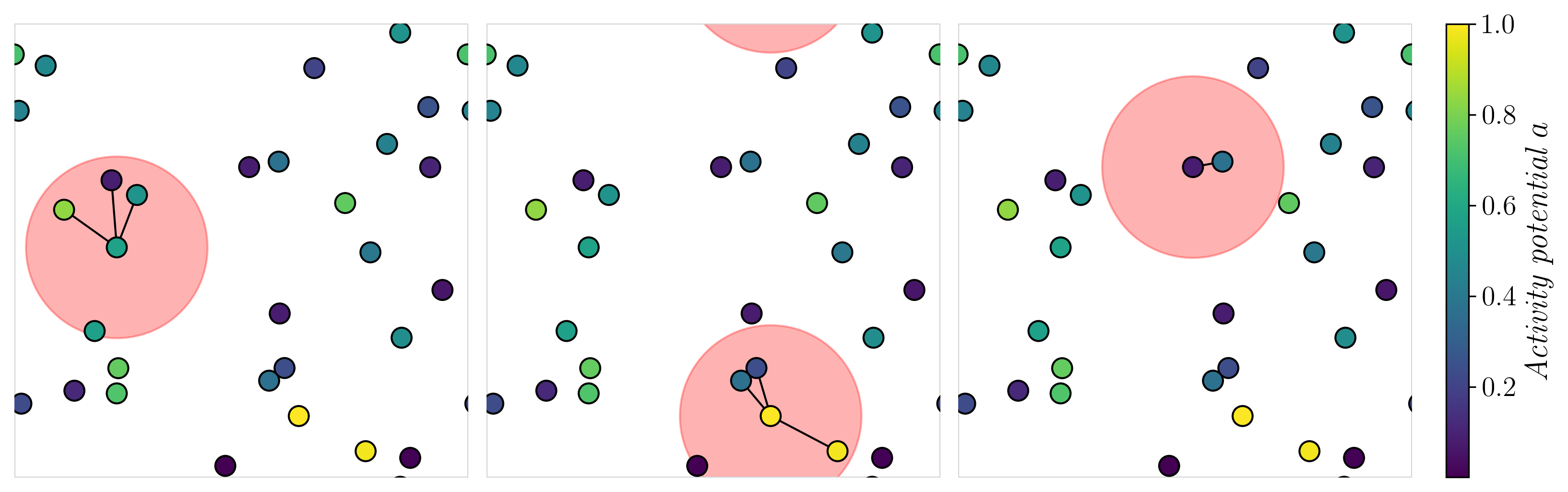}
    \caption{Visualisation of the spatially preferential contact process. $N$ nodes are positioned randomly in a 2D toroidal space, and their activity potential is represented by their colour, such that the lighter a node, the higher its potential. The plot consists of three panels, each showcasing an active node initiating contacts according to the model's mechanism. A red disc of radius $R$ is centred at the selected active node $i$, the $m$ nodes for contacts are sampled from the nodes $j$ within a cut-off distance $R$ with probability $p_{ij}$ inversely proportional to their distance to the active node $i$. Note that if there are fewer than $m$ nodes within the cut-off distance, then only $m_{true}=min\{m, m_{reachable}\}$ contacts occur. For the simulation above, the parameter values were $N=30, R=0.2, m=3$.}
    \label{fig:linkformation}
\end{figure*}

In the following, we introduce the spatial activity-driven model, defined in discrete time. We characterise the resulting series of instantaneous networks by aggregating over an arbitrary time interval. In the following, a \emph{contact} denotes an instantaneous interaction between a pair of nodes, and a \emph{link} exists between a pair nodes in the aggregated network if at least one contact  has occurred between them during the aggregation.\\

\textbf{Setup and Initialisation}\\

First, we initialise our system by assigning an activity potential $a_i,  i= 0,...,N-1$ to each of the $N$ nodes according to an arbitrary activity distribution. Since we would like to model social interactions, it is reasonable to choose a heterogeneous distribution such that there are some nodes with extremely high activity rates. We draw reals from a power law distribution with exponent 10, and then normalise it by the highest value to obtain a heterogeneous distribution with relatively low mean and a handful of highly active nodes. Then we embed the nodes in space by assigning a random position to each node in a 2-dimensional toroidal space. This step lays out the ground for the spatially preferential link formation mechanism, where the distance between nodes will determine the probability of a link appearing between them. As a distance measure, we employ 2D toroidal ("wrap-around") distance, which for two nodes with positions $\mathbf{x}_1=(x_{11},x_{12}), \mathbf{x}_2=(x_{21},x_{22})$ is defined as

\begin{equation}
    d(\mathbf{x}_1, \mathbf{x}_2) = \sqrt{ \sum_{i=1}^{2} \left( \min\left( |x_{1i} - x_{2i}|,\ 1 - |x_{1i} - x_{2i}| \right) \right)^2. }
    \label{toroidalDistance}
\end{equation}

\textbf{Network generation}\\

Let us denote the network at time $t \in [t_0=0,T]$ by $G_t$. Then, the network is generated in discrete steps throughout the time interval $[t_0=0,T]$ as follows:\\

At each time step $t \in [t_0=0,T]$:

\begin{enumerate}
    \item $G_t$ is initialised as $N$ disconnected nodes.
    \item Then, each node may or may not become active according to its own assigned activity potential.
    \item Each active node can initiate $m$ contacts to be formed with $m$ other nodes, which need not be active at the time. The $m$ nodes get selected according to a distance kernel so that for active node $i$, potential neighbours $j$ are assigned a contact probability inversely proportional to their distance to the active node, i.e., to $d(i,j)$, where $d$ is the 2D toroidal distance. This means that the closer a node $j$ is to the active node $i$, the higher the probability that $j$ will be chosen by $i$ to form a contact with.
    \item At the next time step $t'=t+\Delta t$ (for simplicity, we use $\Delta t = 1$), the process returns to step 1, reverting the network to $N$ disconnected nodes. The spatial location and activity potential of nodes do not change.
    \item The aggregated network $G$ at time $T$ is the union of all networks $G_t$ up to that time, i.e. $G = \bigcup_{t \in \{0,...,T\}} G_t$. The weight $w_{ij}$ of a link is defined as the total number of contacts between the nodes $i$ and $j$.
\end{enumerate}

For simplicity, we use a linear kernel for the contact probability with a cut-off radius $R$ such that if $d(i,j) > R$, where $d$ is the 2D toroidal distance,  the probability of contact forming between these two nodes is zero, $p_{ij}=0$. Introducing a cut-off distance is not strictly necessary, but it allows us to avoid computational issues that may arise in the calculation of contact probabilities $p_{ij}$. Since the nodes are sprinkled randomly in space, some nodes may have fewer than $m$ potential neighbours within their interaction range $R$, and therefore the number of contacts formed is always $m_{true}=\min\{m, m_{reachable}\}$.
The probability kernel is thus defined as 

\begin{equation}
\begin{aligned}
p_{ij} = 
\begin{cases}
\displaystyle \frac{1 - \frac{d(i,j)}{R}}{\sum_{k: d(i,k) \leq R} \left(1 - \frac{d(i,k)}{R}\right)}  & \text{if } d(i,j) \le R \\[10pt]
0 & \text{if } d(i,j) > R
\label{probkernel}
\end{cases}
\end{aligned}
\end{equation}

with the normalisation constant

$$
\sum_{k: d(i,k) \leq R} \left(1 - \frac{d(i,k)}{R}\right) := Z_i.
$$

This probability kernel provides a meaningful way to express spatially preferential contacts within the network. It is, of course, possible to use an exponentially decaying or any other monotonically decreasing spatial kernel as an alternative and derive similar results. The spatially preferential contact  process is visualised in Figure \ref{fig:linkformation}.

\section{Results}

\subsection{Properties of the Spatiotemporal Activity-driven Network}

Let us now analytically describe the properties of the aggregated network $G$ generated as detailed above. We start with a node-level (ego-net) perspective, which addresses the structure around individual nodes.
We formulate the analytical expressions for link weight and triangle weight distributions of the aggregated network and demonstrate good agreement with simulation results. Then we show that the simulated aggregated networks display characteristic traits of social networks such as strong and weak ties, high-weight triangles, and high clustering.\\

\textbf{Ego-Net Link Weight Distribution}\\

Recall the definition of contact probability $p_{ij}$ from Equation \ref{probkernel}, and the uniformly random distribution of nodes in 2D toroidal space. Using this kernel, we can now derive the distribution of accumulated link weights between a focal node $i$ and all other nodes $j$ over time $T$, in other words, the ego-net link weight distribution of node $i$ in the aggregated network.

Since at each time step node $i$ becomes active with probability $a_i$ and, if active, initiates $m$ contacts, the probability that node $j$ receives a contact from $i$ at time $t$ is

\begin{equation}
P(i \to j \text{ at } t) = a_i \cdot m \cdot p_{ij}.
\end{equation}

Similarly, the probability that node $i$ receives a contact from node $j$ is 
\begin{equation}
P(j \to i \text{ at } t) = a_j \cdot m \cdot p_{ij}.
\end{equation}

Assuming independence across time steps, the expected accumulated link weight over $T$ time steps is

\begin{equation}
\begin{aligned}
w_{ij}(T) = \mathbb{E}[\text{\# contacts } i \to j, j \to i] \\ = T \cdot (a_i +a_j) \cdot m \cdot p_{ij}.
\label{eq:expected_weight}
\end{aligned}
\end{equation}

Let us make an approximation by treating the density of nodes around $i$ as spatially uniform within the cut-off radius $R$. Let $\rho$ denote the node density in space, i.e., the number of nodes per unit area. Then $\rho = N$ since the torus is a unit square. It follows that the expected number of nodes within distance $r \le R$ is

\begin{equation}
n_i(R) = \rho \pi R^2 = N \pi R^2.
\end{equation}

Now, to compute the expected link weight to nodes at distance $r \le R$, we consider the probability density of distances to nodes under a uniform distribution. The probability that a node lies in the annulus between $r$ and $r+dr$ is

\begin{equation}
dP = \frac{\rho2\pi  r \, dr}{n_i(R)} = \frac{2\pi r \, dr}{\pi R^2} = \frac{2r \, dr}{R^2}.
\end{equation}

Therefore, the expected ego-net link weight distribution for node $i$ is given by the function

\begin{equation}
P(w) = \left| \frac{dr}{dw} \right|  \frac{2r(w)}{R^2}.
\label{eq:weight_distribution}
\end{equation}

On the other hand, the expected weight of the link to a node at distance $r \le R$ can be formulated as

\begin{equation}
w(r) = \frac{T (a_i +a_j) m \left(1 - \frac{r}{R} \right)}{\mathbb{E}[Z_i]},
\end{equation}
 which can be reordered to express $r$ in terms of $w$ as

\begin{equation}
    r(w) = R \left(1 - \frac{w \cdot \mathbb{E}[Z_i]}{T (a_i +a_j) m} \right).
\end{equation}

Differentiating gives

\begin{equation}
\frac{dr}{dw} = - \frac{R \cdot \mathbb{E}[Z_i]}{T (a_i +a_j) m}.
\end{equation}

Plugging this back to Equation \ref{eq:weight_distribution} yields

\begin{equation}
\begin{aligned}
P(w) = \left| \frac{dr}{dw} \right|\frac{2r(w)}{R^2} \\
= \frac{R \mathbb{E}[Z_i]}{T (a_i +a_j) m}  \frac{2}{R^2}  R \left(1 - \frac{w  \mathbb{E}[Z_i]}{T (a_i +a_j) m} \right) \\
= \frac{2  \mathbb{E}[Z_i]}{T (a_i +a_j) m}  \left(1 - \frac{w  \mathbb{E}[Z_i]}{T (a_i +a_j) m} \right).
\end{aligned}
\end{equation}

Therefore, the distribution of link weights in the egonet of node $i$ is

\begin{equation}
\scriptstyle{
P_i(w) = 
\begin{cases}
\displaystyle \frac{2 \cdot \mathbb{E}[Z_i]}{T (a_i +a_j) m} \cdot \left(1 - \frac{w \cdot \mathbb{E}[Z_i]}{T (a_i +a_j) m} \right) \\[5pt] \hspace{1cm}\text{if } w \le \frac{T (a_i +a_j) m}{\mathbb{E}[Z_i]} \\[10pt]
0 \hspace{1cm} \text{otherwise}.
\label{eq: likweightdbn}
\end{cases}
}
\end{equation}

The expected value of the normalisation constant $Z_i$ can be expressed as

\begin{equation}
\mathbb{E}[Z_i] = \int_{0}^{R} \left(1 - \frac{r}{R} \right) 2\pi r \rho \, dr.
\end{equation}

For numerical analysis and comparison with simulated data, we take $\rho = N$, and solving the integral yields

\begin{equation}
\begin{aligned}
\mathbb{E}[Z_i] = 2\pi \rho \int_0^R \left(r - \frac{r^2}{R} \right) dr 
= 2\pi \rho \left[ \frac{r^2}{2} - \frac{r^3}{3R} \right]_0^R \\
 = \frac{\pi \rho R^2}{3} \approx \frac{\pi N R^2}{3}.
\end{aligned}
\end{equation}

Now that we have the expected value of the normalisation constant, we have arrived at a formula for $P_i(w)$ that can be easily implemented computationally:

\begin{equation}
\scriptstyle{
P_i(w) \approx
\begin{cases}
\displaystyle \frac{2 \pi N R^2}{3T (a_i +a_j) m}  \left(1 - \frac{w \pi N R^2}{3T (a_i +a_j)m} \right) \\[5pt] \hspace{1cm} \text{if } w \le \frac{3T (a_i +a_j) m}{\pi N R^2} \\[10pt]
0 \hspace{1cm} \text{otherwise.}
\end{cases}
}
\end{equation}

\begin{figure}
    \centering
\includegraphics[width=1\linewidth]{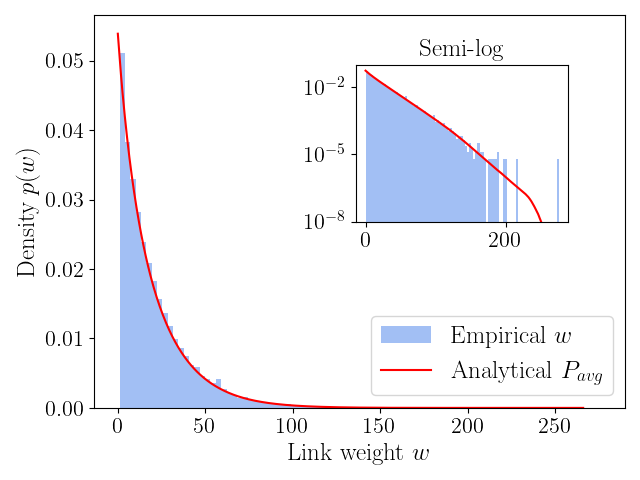}
    \caption{Comparison of theoretical $P_{avg}(w)$ and empirical (simulated) link weight distributions of the aggregated network over $T=2000$ time steps. Further simulation parameters were $N=10^3, R=0.2, m=3$. The analytical result is supported up to the highest observed link weight in the simulation data.}
    \label{fig:analyticallinkweight}
\end{figure}

To analytically approximate the network-level distribution of link weights in the simulated network, we evaluate the node-wise ego-net expression derived above. For each node $i$, the expected total number of its outgoing and incoming contacts is distributed according to a piecewise linear density defined over the interval $[0, Ta_im/Z_i]$, where $a_i$
is the node's activity rate, 
$m$ the number of contact attempts per time step, $T$ the simulation duration, and $Z_i$ the kernel normalisation factor accounting for local neighbour density. The resulting probability density function for node $i$ is given by Equation \ref{eq: likweightdbn}. To account for heterogeneity across the network, we compute this distribution for all $N$ nodes, and average them to obtain the population-level prediction of link weight distribution $P_{avg}(w)$. The resulting analytical distribution $P_{avg}(w)$ is then compared against the empirically observed histogram of total link weights aggregated over the entire simulation, as shown in Figure \ref{fig:analyticallinkweight}.\\

Based on the analytical results and the well-agreeing simulations, we conclude that the spatiotemporal network exhibits an exponential link weight distribution: both strong and weak ties exist, but the tail of the distribution is thin. This is meaningful for social networks, where tie strength is subject to various constraints from the available time to cognitive limits~\cite{Powell2012,Miritello2013}, and in line with empirical results on social networks~\cite{Onnela2007NJP} where comparable distributions have been observed. Furthermore, as with empirical datasets, the exact link weight values depend on the length of the observation window ($T$)~\cite{Krings2012}.
\\
\newpage
\textbf{Triangle Weight Distribution}\\

\begin{figure}
    \centering
    \includegraphics[width=1\linewidth]{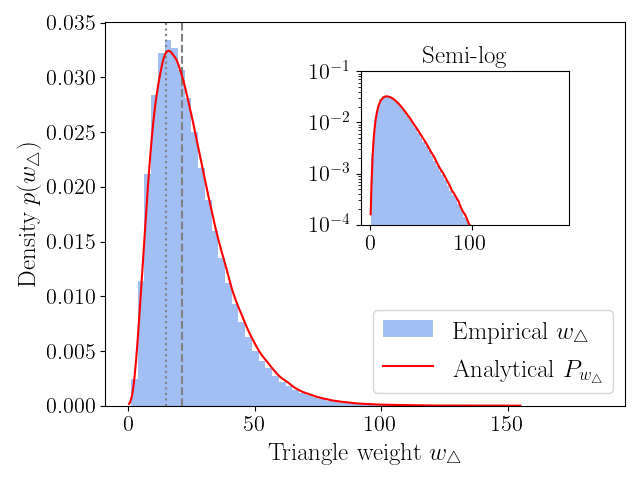}
    \caption{Analytical approximation $P_{w_{\triangle}}$ of the empirical triangle weight distribution. Simulation parameters were $N=10^3, T=2000, R=0.2, m=3$. The approximation was obtained by a Gaussian kernel density estimate of the expected triangle weights (Equation \ref{eq: exp_trigweight}) calculated for each triangle of the aggregated network. The grey dotted and dashed lines indicate the median and mean link weights of the network, respectively. $74\%$ of triangle weights are higher than the median link weight ($54\%$ are higher than the mean link weight, but note the exponential distribution).
    }
    \label{fig:trigw_analytical}
\end{figure}

\begin{figure*}[t]
    \centering
    \begin{subfigure}[b]{0.3\textwidth}
        \centering
        \includegraphics[width=\linewidth]{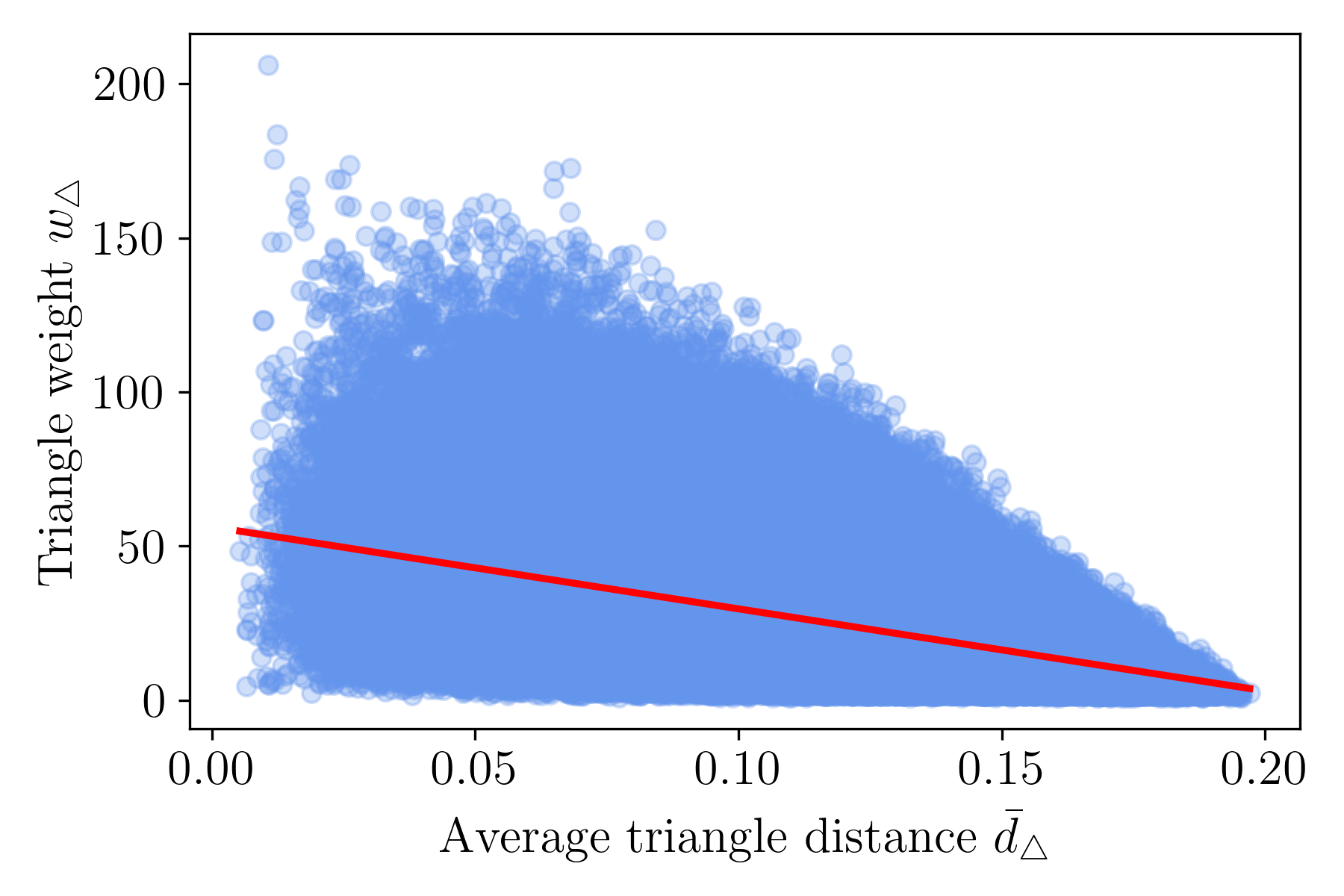}
        \caption{Moderate negative correlation between average triangle distance $\bar{d}_{\triangle}=\frac{1}{3}(d_{ij}+d_{jk}+d_{ki})$ and triangle weight $w_{\triangle}$ for simulated network.}
        \label{fig:correl_dist_weight}
    \end{subfigure}
    \hfill
    \begin{subfigure}[b]{0.3\textwidth}
        \centering
        \includegraphics[width=\linewidth, trim={150 0 100 50}, clip]{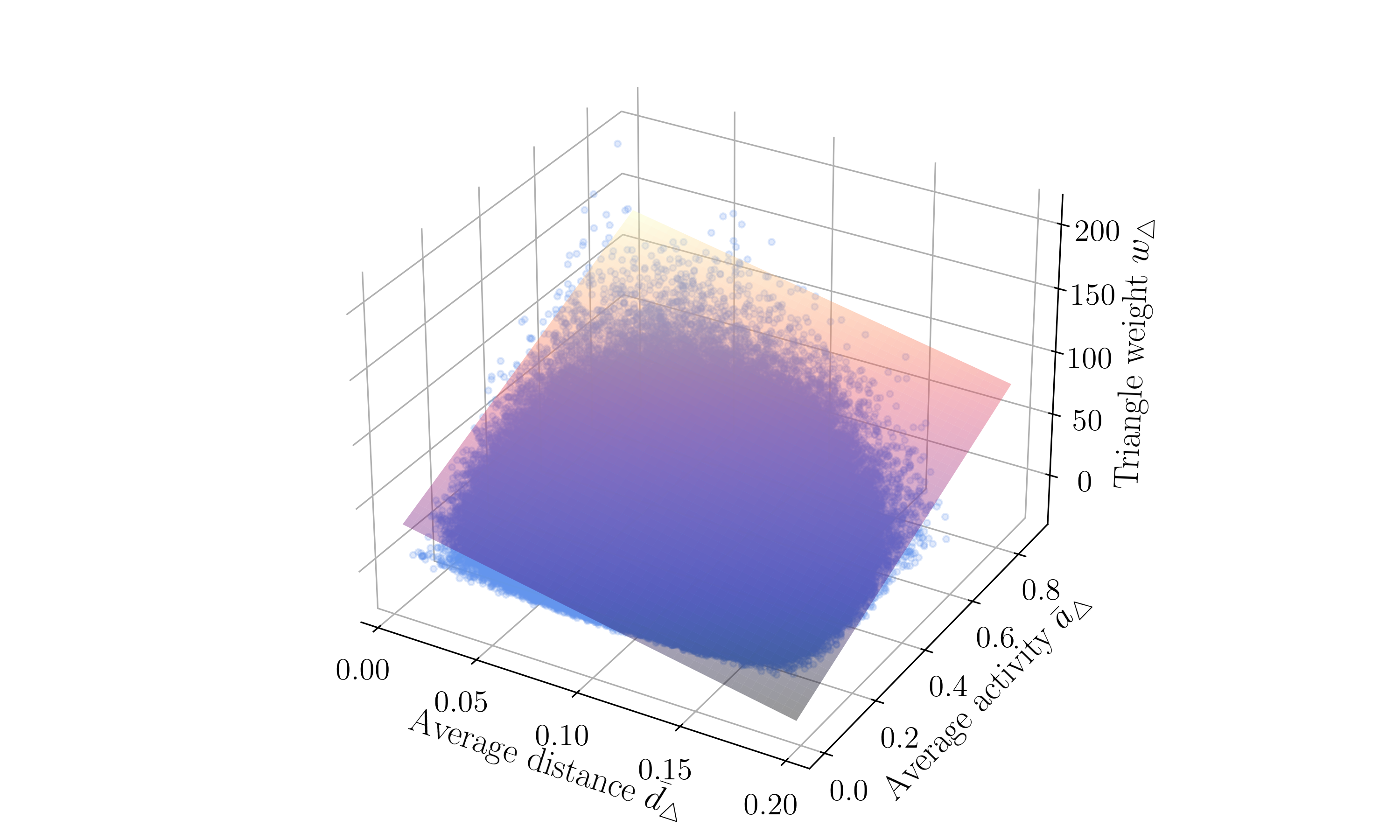}
        \caption{Linear regression surface for triangle weight $w_{\triangle}$, average triangle distance $\bar{d}_{\triangle}$ and average triangle activity $\bar{a}_{\triangle}$ in simulated network.}
        \label{fig:regression}
    \end{subfigure}
    \hfill
    \begin{subfigure}[b]{0.3\textwidth}
        \centering
        \includegraphics[width=\linewidth]{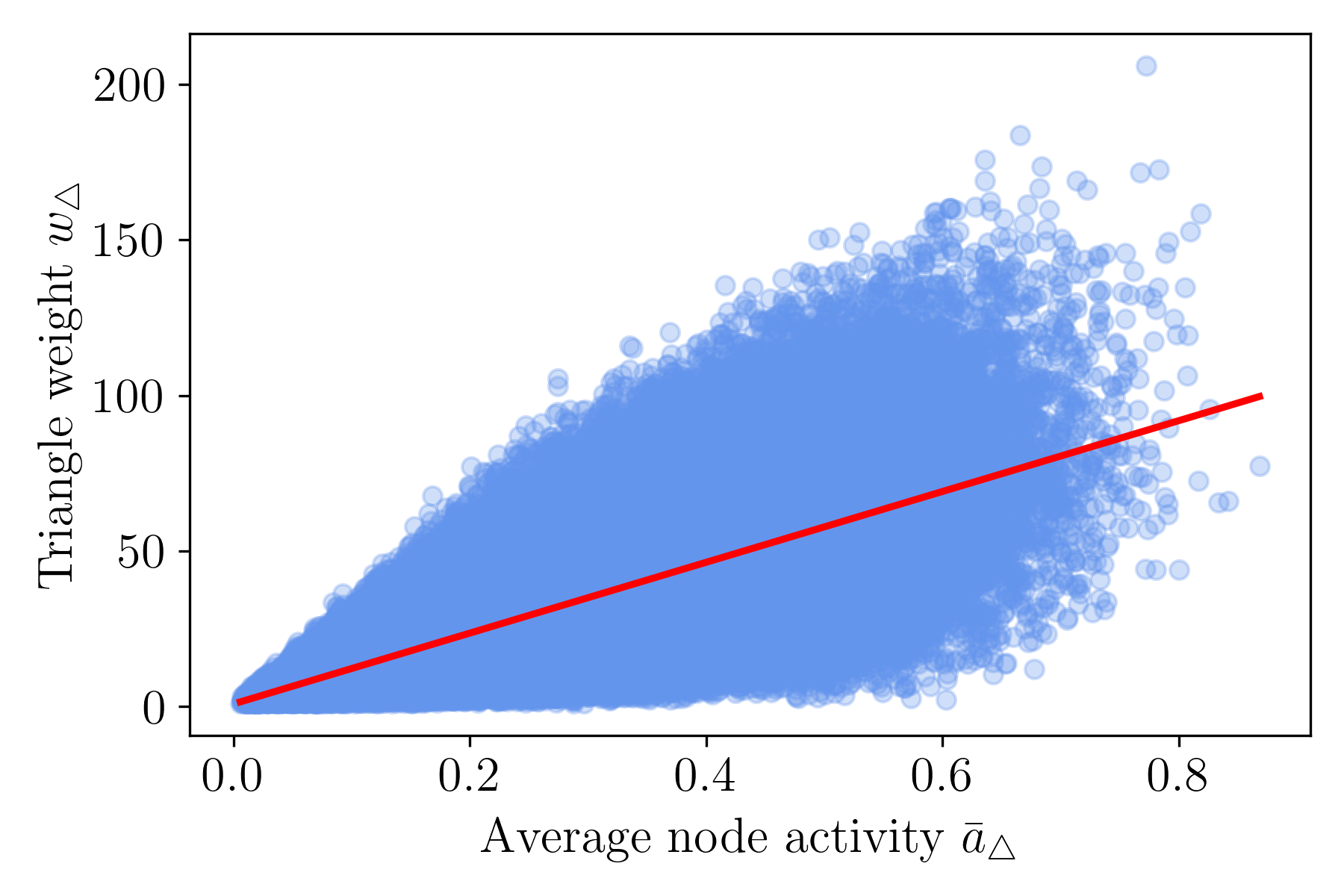}
        \caption{Strong positive correlation between average node activity in triangles $\bar{a}_{\triangle}$ and triangle weight $w_{\triangle}$ for simulated network.}
        \label{fig:correl_activity_weight}
    \end{subfigure}

    \caption{Correlation (linear regression) results for spatiality and activity of triangles with triangle weight. Blue dots denote triangles in the simulated aggregated network, with parameters $N=10^3, T=2000, R=0.2, m=3$. Similar results were obtained for different $R$ and $m$ values. The regression surface shows moderate negative correlation (Pearson, Spearman $< -0.4$) with average triangle distance $\bar{d_{\triangle}}$, and strong positive correlation (Pearson, Spearman $> 0.7$) with average activity of nodes $\bar{a}_{\triangle}$, verifying our hypothesis.}
    \label{fig:trigweight}
\end{figure*}

In the aggregated network, we define the weight of a triangle to be the mean of its edge weights. We then consider the weight distribution of triangles in the aggregated network with particular focus on the spatial relationship of node triplets as well as their activity level. For a high-weight link to form, multiple contacts over time are needed, and this is more likely if the two nodes are highly active and spatially close. Therefore, directly following from the way the network is generated, one can expect that the closer the three nodes are to each other and the higher their activity potential, the higher the weight of their triangle. This hypothesis has been tested and verified by running simulations with consistent outcomes, and an example case is shown along with regression results in Figure \ref{fig:trigweight}. Below, we derive the analytical expression for triangle weights and demonstrate its agreement with empirical data.

We define the weight of a triangle formed by nodes $i,j,k$ as the mean of the weights of its three edges,

\begin{equation}
    w_{\triangle ijk} = \frac{1}{3}(w_{ij} + w_{jk} + w_{ki}).
\end{equation}

Following the above derivations, the expected weight of an edge between $i$ and $j$ over $T$ time steps is

\begin{equation}
\begin{aligned}
    w_{ij} = T \cdot (a_i +a_j) \cdot m \cdot p_{ij} = \frac{T (a_i +a_j) m}{\mathbb{E}[Z_i]} \left(1 - \frac{d_{ij}}{R} \right)\\ \scriptstyle{\text{for } d_{ij} \le R.}
\end{aligned}
\end{equation}

Recall that a contact $(u,v)$ may be initiated by either node $u$ or $v$. Therefore, we can compute the expected triangle weight as follows:

\begin{equation}
\begin{aligned}
w_{\triangle ijk}
&=\frac{Tm}{3} \Bigg( 
    \frac{a_i}{Z_i} \left(2 - \frac{d_{ij}+d_{ki}}{R} \right) \\ &
  + \frac{a_j}{Z_j} \left(2 - \frac{d_{jk}+d_{ij}}{R} \right)  + \frac{a_k}{Z_k} \left(2 - \frac{d_{ki}+d_{jk}}{R} \right) 
\Bigg).
\end{aligned}
\label{eq:triangle-weight}
\end{equation}

Now we approximate all normalisation constants $Z$ using the same uniform density assumption within the cut-off radius $R$,

\begin{equation}
    Z \approx \mathbb{E}[Z] = \int_{0}^{R} \left(1 - \frac{r}{R} \right) 2\pi r \rho \, dr = \frac{\pi N R^2}{3}
\end{equation}
which in turn simplifies the expected triangle weight to

\begin{equation}
\begin{aligned}
    w_{\triangle ijk} &\approx \frac{T m}{\pi N R^2} \cdot \Bigg(  a_i \left(2 - \frac{d_{ij}+d_{ki}}{R} \right)\\
&\quad\;\; + a_j \left(2 - \frac{d_{jk}+d_{ij}}{R} \right) + a_k \left(2 - \frac{d_{ki}+d_{jk}}{R} \right) \Bigg).
\end{aligned}
\label{eq: exp_trigweight}
\end{equation}

Finally, this can be reordered as 
\begin{equation}
\begin{aligned}
    w_{\triangle ijk} \approx \frac{Tm}{\pi N R^2} \Biggl[ 2(a_i+a_j+a_k)- \\
    \left( a_i \frac{d_{ij}+d_{ik}}{R}+  a_j \frac{d_{ij}+d_{jk}}{R} + a_k \frac{d_{jk}+d_{ik}}{R}\right) \Biggr].
\end{aligned}
\label{eq: trigw_reordered}
\end{equation}

To analytically approximate the empirical triangle weight distribution, we apply the above expression from Equation \ref{eq: exp_trigweight} and proceed by sampling all triangles from the simulated network, evaluating their expected weights, and constructing an analytical distribution by fitting a Gaussian kernel density estimate to them. The resulting curve closely matches the empirical one, as shown in Figure \ref{fig:trigw_analytical}. Notice that over $70\%$ of the triangles are heavier than the median link weight.

Note that the reordered expression (Equation \ref{eq: trigw_reordered}) clearly underpins the empirically observed moderate negative correlation with average distance of node triplets in triangles, as well as the strong positive correlation with the average activity potential of such node triplets, all in agreement with the results in  Figure \ref{fig:trigweight}.\\

\textbf{Evolution of Link Density}\\

\begin{figure}
    \centering
    \includegraphics[width=1\linewidth]{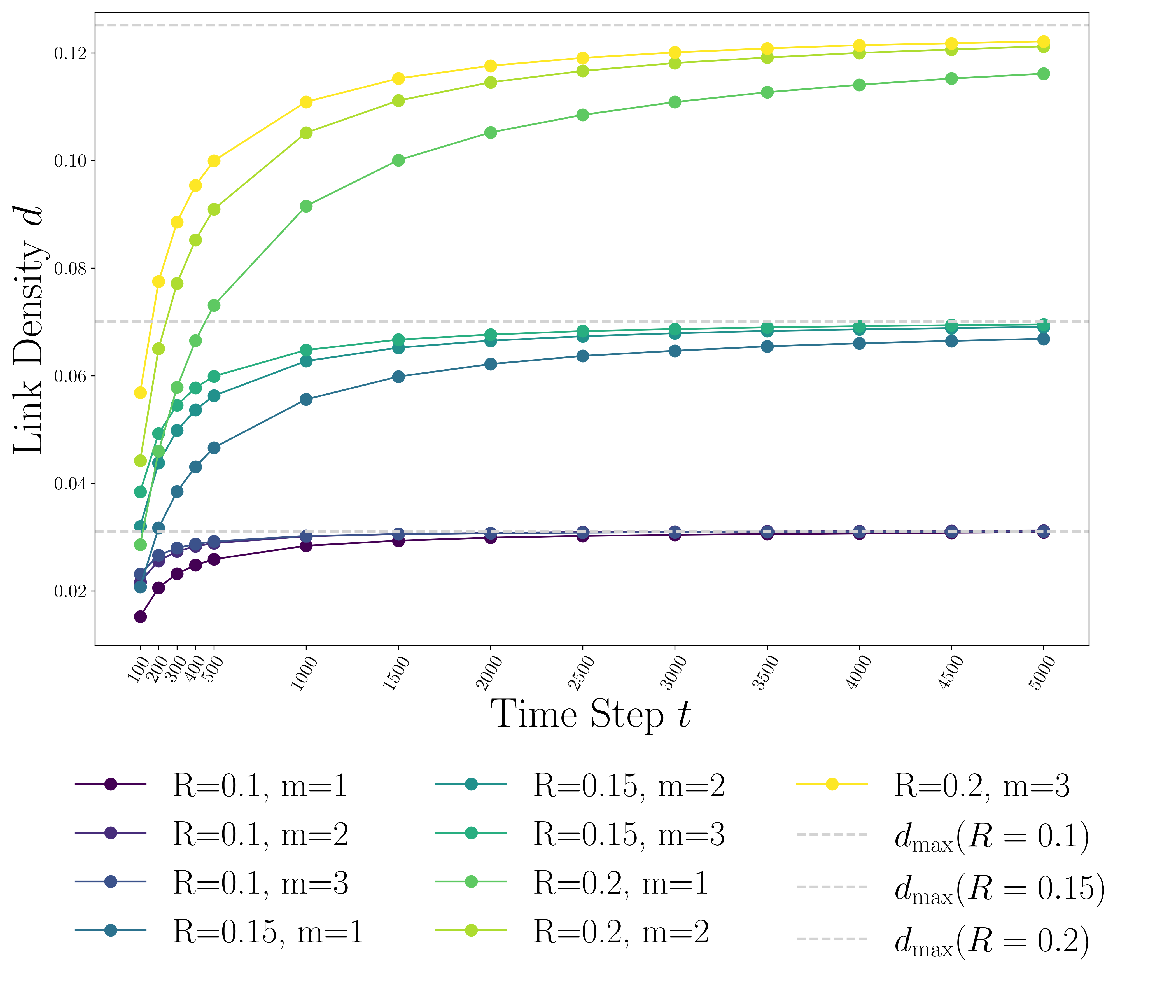}
    \caption{Evolution of link density for different $R,m$ parameter values on networks of $N=10^3$ nodes. Links are aggregated and density values are evaluated at checkpoints in time from $t_0=100$ up to $T=5000$. $d_{max}(R)$ values are marked with grey dashed lines.}
    \label{fig:link_density}
\end{figure}

The link density of a network $G$ of $N$ nodes with edge set $E$ is defined as

\begin{equation}
    d = \frac{2|E|}{N(N-1)}.
\end{equation}

Due to model construction, link density increases throughout the simulation of a network, gradually saturating towards an intrinsic value $d_{max}(R)$ constrained by the interaction radius $R$. The larger $R$ is, the higher this final density value can be, but the slower the convergence of the system. For a chosen $R$, convergence can be sped up by using higher values of $m$.

We can analytically obtain an approximation for this maximum density $d_{max}(R)$  under the assumption that the nodes are uniformly positioned in the unit toroidal space and the network size $N$ is large. Taking the limit $T \rightarrow \infty$, we will have all possible edges present in the aggregated network, meaning that the maximum attainable link density equals the fraction of node pairs that lie within radius $R$:

\begin{equation}
    d_{max}= \frac{2|E_{max}|}{N(N-1)} \approx \pi R^2.
\end{equation}

This holds as long as $R$ is sufficiently small relative to the size of the toroidal space, as in the large $R$ limit, we eventually get $d_{max} \approx 1$ as $T \rightarrow \infty$. Note that this agrees with the behaviour of the simple activity-driven model, for which $d_a \rightarrow 1$ as $T \rightarrow \infty$. A comparative empirical analysis on the effect of $R$ and $m$ on the evolution of $d$ and convergence to $d_{max}$ is shown in Figure \ref{fig:link_density}.

\subsection{Comparison with the Non-Spatial Activity-Driven Model}

To identify the effects of spatiality on the resulting aggregated networks, we compare them with networks generated by the original activity-driven model, contrasting link and triangle weight distributions as well as monitoring the evolution of the clustering coefficient.\\

\textbf{Link and Triangle Weight Distributions}\\

Recall that in the original activity-driven model, active nodes choose their contacts uniformly at random from all $N-1$ available nodes of the network. Conversely, in the spatiotemporal model, contacts are chosen preferentially according to distance, and only nodes that lie within the active node's interaction disc $R$ are available. This difference in the generating mechanisms results in distinct link and triangle weight distributions for the models.

In particular, thanks to uniform link selection probability and access to all nodes, the simple activity-driven model rarely repeats contacts between a pair of nodes, resulting in a network composed of weak links whose number grows with time. The larger the network, the less likely a repeated contact is per unit time. 
On the other hand, due to the limiting radius $R$ and distance-based preferential contact choice, the spatiotemporal model produces repeated contacts frequently and in proportion to the spatial distance spanned by the link, resulting in a highly heterogeneous link weight distribution with both strong and weak ties. This is a significant improvement towards modelling human social interactions since they exhibit similarly heterogeneous link weight properties.\\

\begin{figure}
    \centering
    \includegraphics[width=1\linewidth]{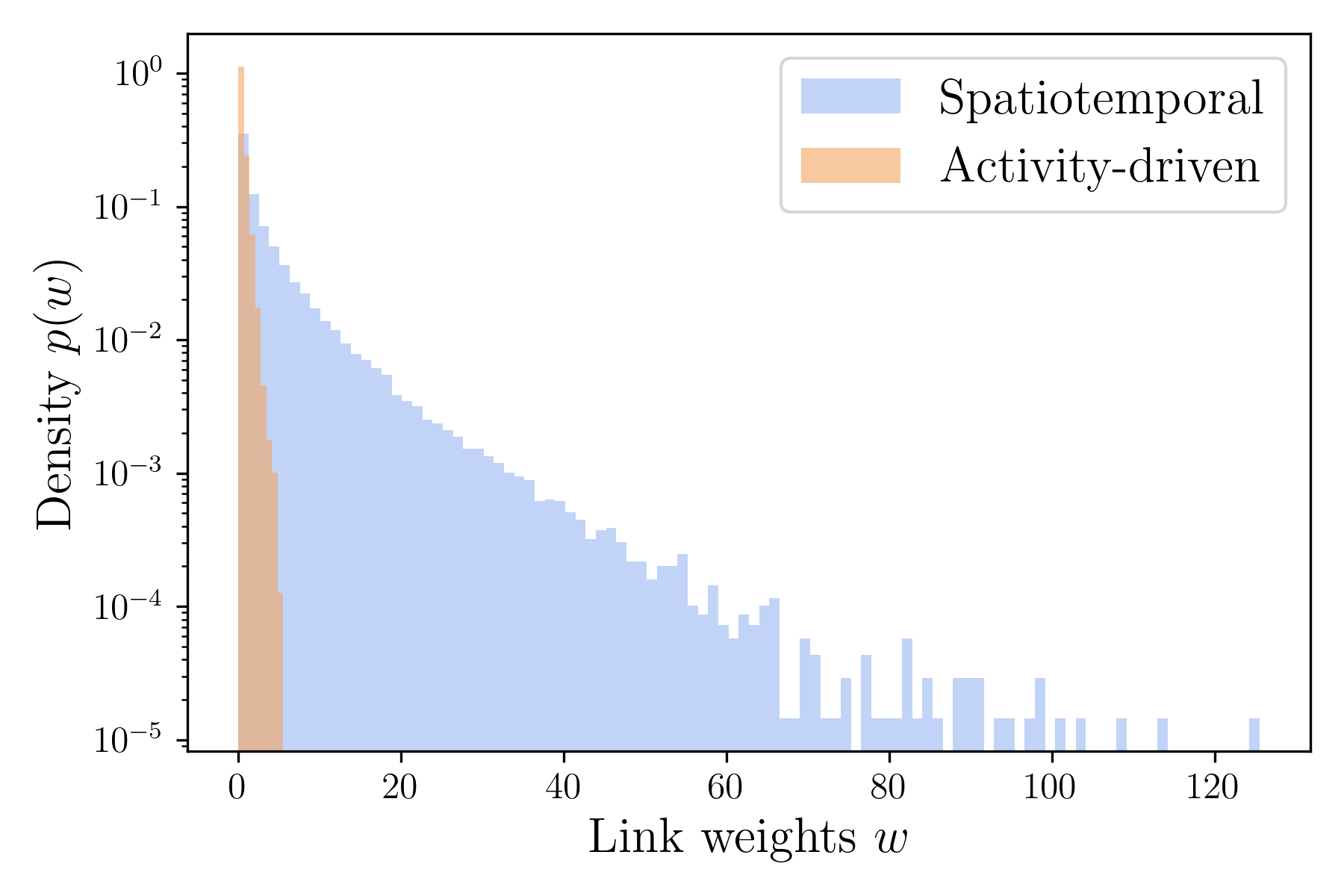}
    \caption{Comparison of empirical link weight distributions of spatiotemporal and simple activity-driven networks. Parameter values ($N=10^3, T=1000, R=0.15, m=3$) as well as activity potential of nodes were kept the same for the simulations.}
    \label{fig:linkw_compare}
\end{figure}

\begin{figure}
    \centering
    \includegraphics[width=1\linewidth]{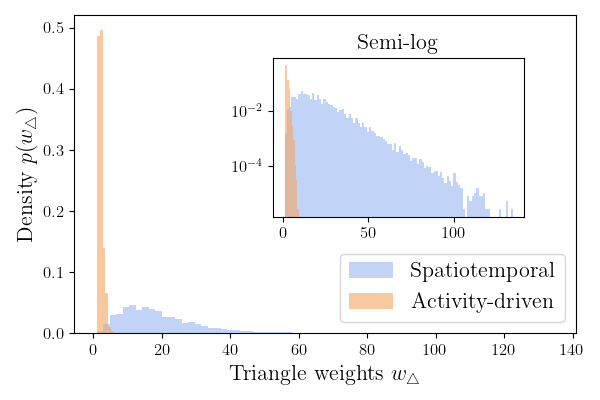}
    \caption{Comparison of empirical triangle weight distributions for spatiotemporal and simple activity-driven network models. Parameter values ($N=10^3, T=1000, R=0.15, m=3$) as well as activity potential of nodes were kept the same for the simulations.}
    \label{fig:trigw_compare}
\end{figure}

These differences carry over to triangle weights where they are amplified: the distribution for the activity-driven model is even narrower with very light triangles, while the spatiotemporal model produces a highly heterogeneous distribution featuring some heavy triangles. 
Figures \ref{fig:linkw_compare} and \ref{fig:trigw_compare} confirm these results with empirical simulated data.\\

%\vspace{0.5cm}
\textbf{Clustering Coefficient}\\

Several studies have shown that interaction networks are highly clustered ~\cite{wasserman1994social} as individuals' social networks are densely interconnected. We examine the effect of space and time on the average local clustering coefficient to determine how well the proposed model captures this structural phenomenon; a reasonably high clustering coefficient can be expected in the aggregated networks due to the known effect of spatiality (e.g.,~\cite{Barthelemy2003}).The clustering coefficient of a node describes what fraction of the possible triangles through it are actually present in the network:

\begin{equation}
    c_i = \frac{\# \triangle_{i,\_,\_} }{k_i(1-k_i)}
\end{equation}
where the numerator refers to the number of triangles containing node $i$, and $k_i$ is this node's degree.

\begin{figure}
    \centering
    \includegraphics[width=1\linewidth]{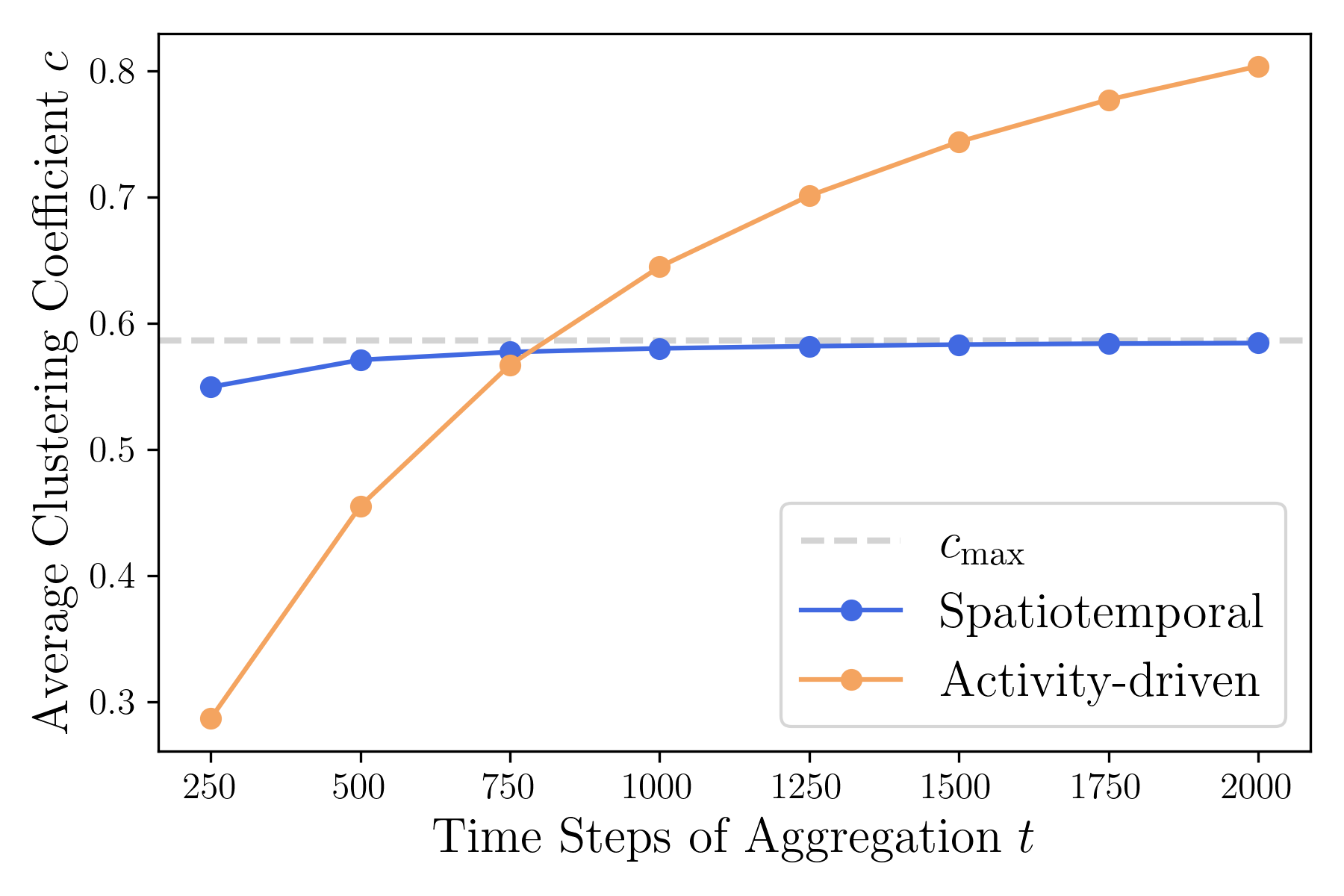}
    \caption{Comparison of the evolution of the average clustering coefficient for spatiotemporal and activity-driven networks. Simulation parameters were $N=10^3, T=2000, R=0.15, m=3$.}
    \label{fig:comparison_clustering}
\end{figure}

\begin{figure*}[t]
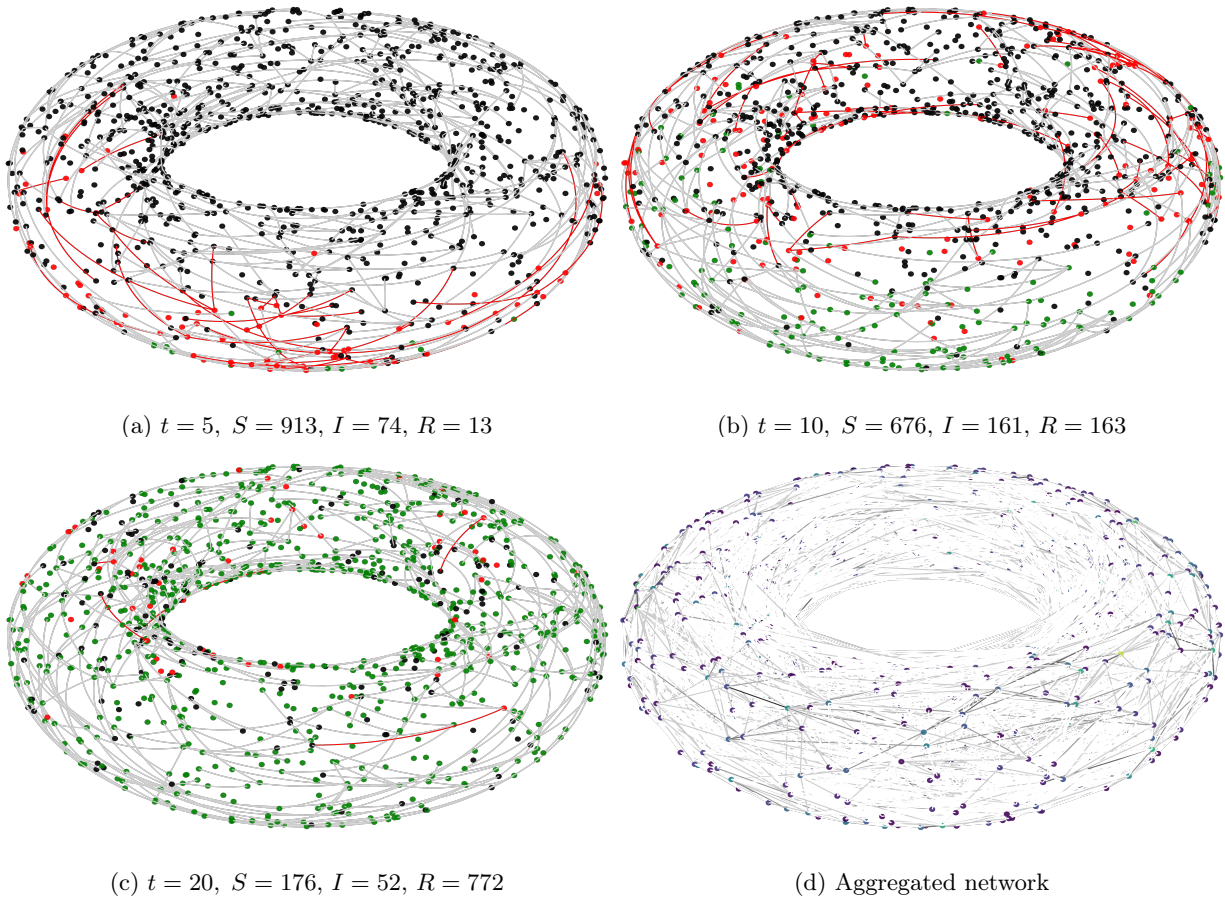

    \centering

    \begin{subfigure}[b]{0.45\textwidth}
        \centering
        \includegraphics[width=\linewidth, trim={30cm 40cm 30cm 100cm}, clip]{figures/Toroidal_epidemic_t5.pdf}
        \caption{$t = 5,\; S = 913,\, I = 74,\, R = 13$}
        \label{fig:toroidal_t5}
    \end{subfigure}
    \begin{subfigure}[b]{0.45\textwidth}
        \centering
        \includegraphics[width=\linewidth, trim={30cm 40cm 30cm 100cm}, clip]{figures/Toroidal_epidemic_t10.pdf}
        \caption{$t = 10,\; S = 676,\, I = 161,\, R = 163$}
        \label{fig:toroidal_t10}
    \end{subfigure}

    \vspace{-0.5em} 
    \begin{subfigure}[b]{0.45\textwidth}
        \centering
        \includegraphics[width=\linewidth, trim={30cm 40cm 30cm 100cm}, clip]{figures/Toroidal_epidemic_t20.pdf}
        \caption{$t = 20,\; S = 176,\, I = 52,\, R = 772$}
        \label{fig:toroidal_t20}
    \end{subfigure}
    \begin{subfigure}[b]{0.45\textwidth}
        \centering
       \includegraphics[width=\linewidth, trim={30cm 40cm 30cm 100cm}, clip]{figures/Toroidal_aggregated_network.pdf}
        \caption{Aggregated network}
        \label{fig:aggregate_network}
    \end{subfigure}

    \caption{SIR spreading process on top of a spatiotemporal network with $N=10^3$ nodes visualised in toroidal space. Simulation parameters were $R=0.2, m=3, \delta=0.9$ and fixed $t_r =3$. The first three panels show $t \in \{5, 10, 20\}$ time instance snapshots of the network with temporal contacts, while the last panel shows the aggregated network at the end of the simulation (27 time steps). On snapshot plots nodes are coloured according to their state with black (S), red (I) and green (R), present contacts are marked with grey and transmitting contacts are highlighted with red. On the aggregated plot nodes are coloured according to their activity potential and the weight of edges is represented by their intensity.}
    \label{fig:toroidal_sir_snapshots}
\end{figure*}

For both network types we aggregate all edges up to time $t$, then evaluate and average this measure $c_i$ for all nodes to get $c$. We found that the purely activity-driven model displays a steadily growing trend, while the spatiotemporal model's clustering always converges to the same value, which is inherent to the model.  
In fact, in the limit of $T \rightarrow \infty$, the activity-driven model reaches $c \rightarrow 1$ since all possible edges will eventually exist, and the system converges to a complete graph.
On the other hand, in the limit of $T \rightarrow \infty$, the spatiotemporal model accumulates all possible edges between pairs of nodes within distance $R$, converging to a random geometric graph (RGG) \cite{dall2002random} with connection radius $R$. In the Appendix, we show analytically that regardless of the values of $R$ and $m$, the average clustering coefficient of the spatiotemporal model converges to the system's inherent value, which is 

\begin{equation}
    c \rightarrow c_{max} = 1- \frac{3 \sqrt{3}}{4 \pi} \approx 0.5865 
\end{equation}

Simulation results support the analysis as illustrated in Figure \ref{fig:comparison_clustering}. The speed of convergence to this unique clustering value depends on the parameters $R$ and $ m$ as well as the spatial distribution of nodes.
In comparison, human interaction networks have been shown to exhibit consistent and relatively high clustering, which agrees with our findings.\\

We conclude that introducing spatiality to the activity-driven framework induces highly heterogeneous link and triangle weight distributions, producing strong and weak ties as well as heavy triangles. The clustering coefficient of the network quickly saturates to a high inherent value determined by the spatial distribution of nodes in the network.

\begin{figure*}
    \centering
    \includegraphics[width=1\linewidth]{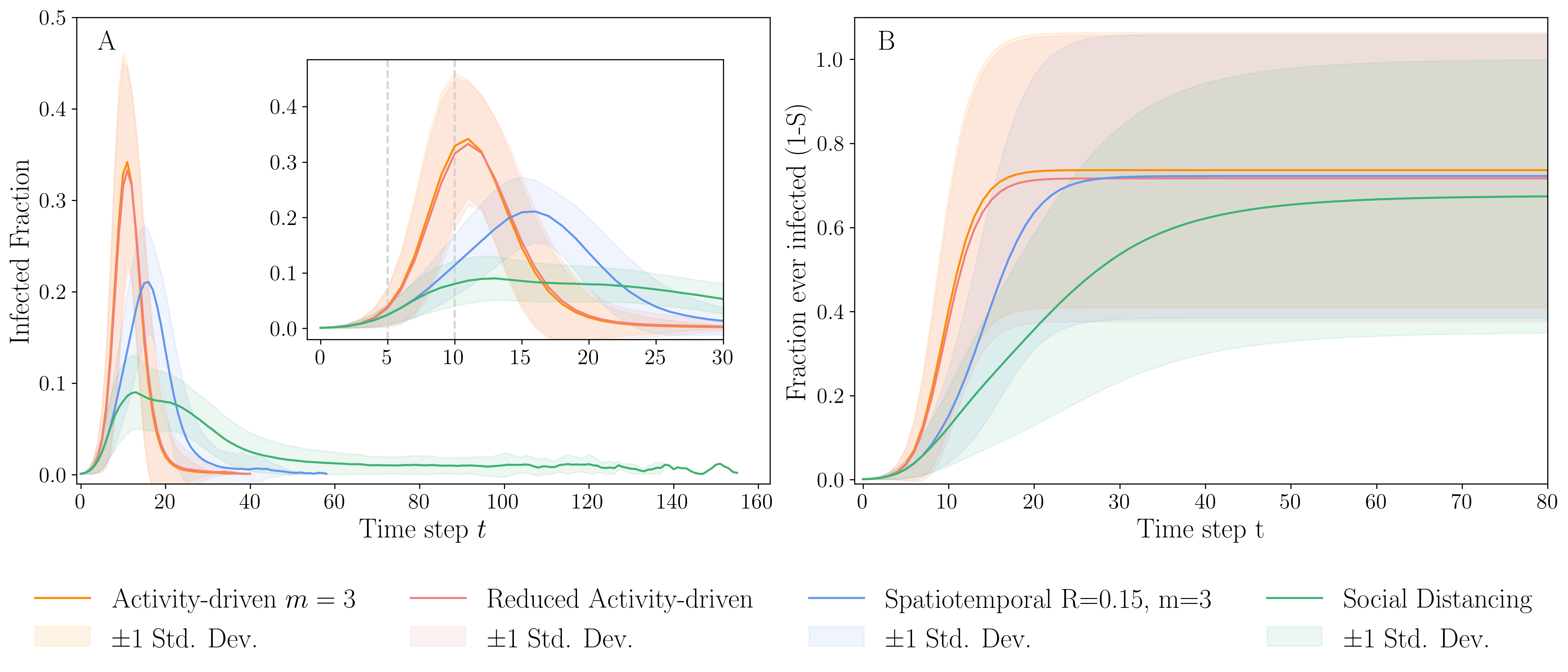}
    \caption{Mean simulated SIR infection curves with zoomed-in inset (panel A) and cumulative mean curves for the fraction of ever infected nodes (panel B). The spatiotemporal and activity-driven models have agreeing number of temporal contacts and their reduced counterparts have $1.5\%$ less contacts than the originals. The reduction is done by random contact removal for the activity-driven setup, while for social distancing it is an induced effect of the shrunken interaction discs. For each of the 4 models a total of 5000 simulations were run (50 configurations of $N=10^3$ nodes, 100 runs on each), and the mean curves are presented with a 1 standard deviation wide band around them. The spatial locations and activity potentials of nodes were kept the same for quadruples of configurations, all configurations had $m=3$, and all SIR simulations had parameters $ \beta=0.9, t_r=3$. For the spatiotemporal model the radii were $R=0.15$, while social distancing was obtained by a 2-step reduction of interaction radii: first to $R=0.1$ at $t_1=5$, then further tightened to $R=0.075$ at $t_2=10$. Intervention times $t_1,t_2$ are marked with grey dashed lines in the inset plot of panel A.}
    \label{fig:infections}
\end{figure*}

\subsection{Applications - SIR Epidemics}

An interesting application for spatiotemporal activity-driven networks is the modelling of spreading processes such as epidemics of infectious diseases or rumours spreading in a social network. The joint spatial and temporal nature, together with the social network-like characteristics make the model well-suited for this application.

Let us now consider a simple SIR spreading process~\cite{hethcote2000mathematics}. In this setup, individual nodes belong to one of the three classes at a time: susceptible (S), infectious (I) or removed (R), the latter containing both recovered and deceased individuals. The system is initialised so that all nodes are in the susceptible state, except for one seed node which is infectious. Then, simultaneously with the network's spatiotemporal link formation dynamics, the spreading process evolves by passing on the infection from one individual to another via temporal links.
Whenever an infected node comes into contact with a susceptible one, the disease may be passed on with transmission probability $\beta \in (0,1]$, which is a tunable parameter. If the transmission is successful, the state of the susceptible node turns into infectious ($I+S \rightarrow 2I$). At the time of infection, infected nodes are assigned a recovery time $t_{r}$ measured in time steps. For simplicity, we assume constant recovery time. When the required time $t_{r}$ has elapsed, the node's state automatically turns from infectious to removed ($I \xrightarrow[]{t_{r}} R$). Once a node enters the removed state, it cannot be infected again as it is considered to be immune. The spreading simulation stops when there are no more infected nodes left in the network. In the following, we will use the values $\beta=0.9, t_r=3$, which are well above the epidemic threshold. Please note that the aim is to demonstrate the effect of spatiality, rather than provide an extensive sweep of the parameter space.

In addition to examining SIR spreading on the spatiotemporal and simple activity-driven networks with equal numbers of events, we also model social distancing by decreasing the interaction radius $R$ in the spatiotemporal setup. This is in line with empirical observations made during the COVID-19 pandemic~\cite{Schlosser20}. As a reference, we create a comparable non-spatial activity-driven counterpart from the activity-driven network by random edge removal. The resulting reduced activity-driven model has the same number of edges as the social distancing spatiotemporal network.
\\

\textbf{The Role of Space}\\

\begin{figure*}
    \centering
    \includegraphics[width=1\linewidth]{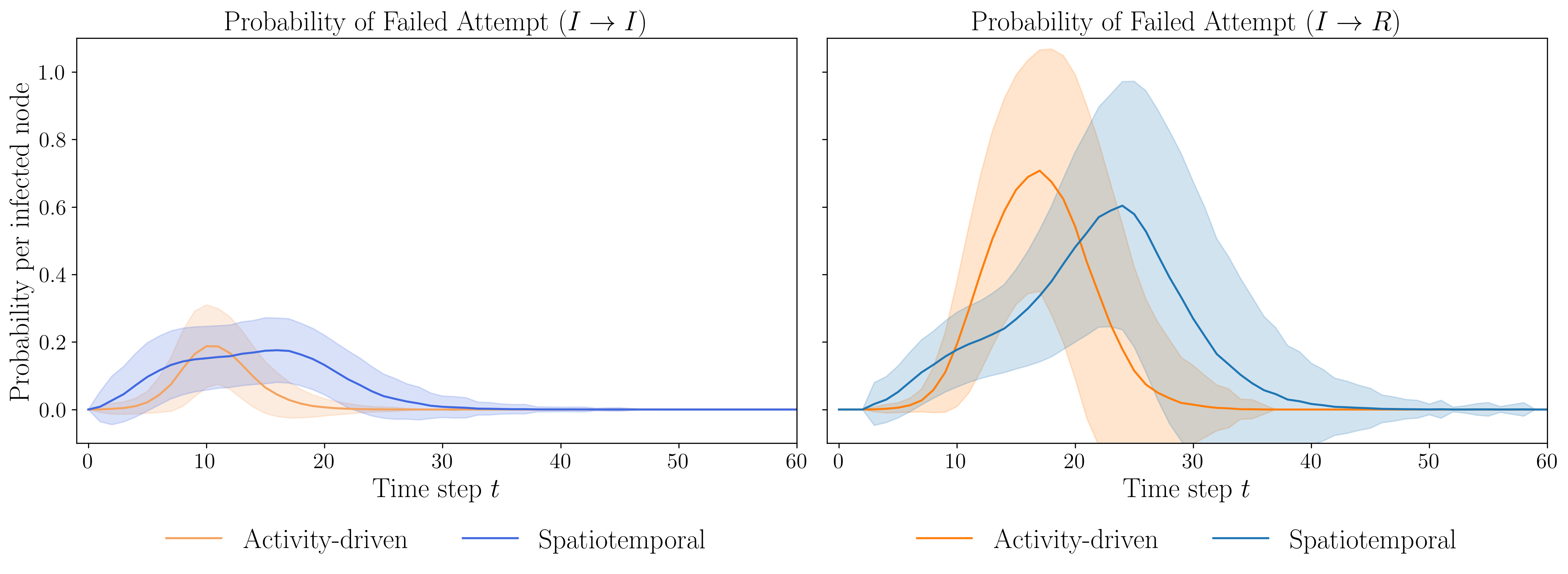}
    \caption{The evolution of mean failed infection probabilities $I \rightarrow I$ and $I \rightarrow R$ per infectious node during SIR simulations. Infectious nodes can only transmit the disease to susceptible (S) ones; other types of contacts do not result in transmission. We call these $I \rightarrow I$, $I \rightarrow R$ events failed infection attempts. For both models, 50 network configurations of $N=10^3$ nodes were used with 100 SIR runs per configuration, yielding 5000 simulations per network type. Further parameters were as before, $\beta = 0.9, t_r = 3$. The bands around the mean curves are 1 standard deviation wide.}
    \label{fig:prob_failed}
\end{figure*}

We examine simulated SIR dynamics on spatiotemporal networks of  $N=10^3$ nodes starting from a randomly chosen seed node, with runs that lasted up to 59 time steps, with up to 27,346 events. One such simulation run is visualised in Figure \ref{fig:toroidal_sir_snapshots}, where nodes are coloured according to their state and transmissive events are highlighted in red. Figure \ref{fig:aggregate_network} shows the aggregated spatiotemporal network up 
to the point of simulation termination, with varying intensities of links indicating their respective heterogeneous weights. It is clear that due to spatial restrictions in link formation---and thus in disease transmission too---the infectious nodes are initially localised to a small region of the network, which then shifts and expands over time. In contrast, there cannot be any spatially localised contagion or expansion in the simple, non-spatial activity-driven model.

More quantitatively, the effect of space is seen in the time series of infections in Figure \ref{fig:infections}.  The evolution of the fraction of infected nodes clearly shows that, as expected, spatiality slows down the spreading process and reduces the epidemic peak height compared to the non-spatial activity-driven model. This is visible in the peak times and heights of infection curves (panel $A$), as well as in the slower cumulative growth in the spatial model (panel $B$).

The role of space is also expected to be visible in the number of events where the contact made by an infectious node cannot transmit the disease due to its neighbour being already infected ($I-I$) or recovered ($I-R$). In a spatial system, spreading can be viewed as the propagation of a spreading front where nodes in the direction of propagation are susceptible, while those behind the front are mostly infectious or recovered. In the original activity-driven model, there is no such localisation as all nodes can be in contact with anyone else. 

Because of these differences, one can expect that the time evolution of the probability of failed contacts, either due to the other node being in the I or R state, would differ for the two models as well.\\ 
 
Figure \ref{fig:prob_failed} illustrates the mean evolution of the probability of failed attempts with 5000 runs across 50 network configurations per model. For the activity-driven model, the probability of failed attempts follows the infection curve of Figure \ref{fig:infections}A. This is because the infectious nodes choose their contacts uniformly randomly from all other nodes of the network. However, in the spatiotemporal model, more interesting behaviour can be observed. At early times, the probability of failed attempts is higher in the spatiotemporal model due to localisation, but it grows more slowly than in the activity-driven setup until an inflection point, which corresponds to the front having circulated around the system. As the radius is $R=0.15$, we can approximate that on average the infection jumps the distance $0.075$ per time step, and in the unit space this corresponds to approximately 13 time steps for going around the system. This roughly matches the results shown in Figure \ref{fig:prob_failed}.\\

\textbf{Spatial Constraints \& Social Distancing}\\

Returning to the infection curves of Figure \ref{fig:infections}, in the social distancing model the interventions introduced at $t_1=5$ and $t_2=10$
have a dramatic effect on the spreading both in terms of speed and peak height, even though social distancing decreases the total number of events by only 1.5$\%$. This is because the interventions effectively increase the geodesic distances between nodes: after the intervention, more intermediate nodes are required in the time-respecting paths between distant nodes. In contrast, in the non-spatial model, the $1.5\%$ reduction in events barely has any effect. 

It is also worth noting that the total number of infections is lower in the case of social distancing than without it in the spatiotemporal model (Fig.~\ref{fig:infections}, panel $B$). This can be attributed to at least two reasons: first, reducing the radius of contacts may give rise to isolated nodes or small components in the spatiotemporal network. Second, as the spreading wave progresses through the network, it may be more likely that individual nodes that escape infection from their nearest neighbours are surrounded and protected by recovered nodes, as there are fewer long-distance links that would allow the disease to re-enter their neighbourhood.

Based on these observations, the proposed spatiotemporal model seems to be well-suited for describing the mechanism and effects of social distancing interventions. Embedding the nodes in latent space, our framework captures the existence of social circles via the interaction radius $R$, with distance-based interaction probabilities ensuring that the heterogeneous contact frequencies within one's acquaintances are also reproduced. The underlying space thus encapsulates social structure, and social distancing effectively makes the social neighbourhoods of individuals shrink to their closest and most frequent contacts. Consequently, the long-range weak ties---which are known to play a crucial role in spreading ~\cite{onnela2007structure,Schlosser20}---are cut, thus hindering spreading across the network, restricting the propagation to short range. This results in, as we have seen, the infection curve flattening out, a much lower peak, and a somewhat lower final cumulative number of infected individuals. These outcomes mirror the empirically observed impact of real-world social distancing measures ~\cite{Schlosser20}.

In conclusion, our model not only highlights the critical importance of the often neglected spatial dimension, but also allows for encoding spatial constraints such as social distancing measures.

\section{Discussion}

To summarise, we have introduced a modelling framework for spatiotemporal networks that is analytically tractable and produces realistic networks without constructing explicit mechanisms to generate their traits. Notably, the constraints imposed by spatiality alone result in network characteristics that resemble those of real-world social networks: high clustering, the existence of strong and weak ties, and high-weight links associated with triangles. Additionally, the model allows us to investigate the role of time and space separately as well as together to reveal their joint effects. Due to these social-network-like attributes and the spatiotemporal nature of the networks, the model is directly applicable to dynamic epidemic modelling on human interaction networks. 

We have demonstrated with simulations how spatiality slows down spreading: while at macro scale this is obviously related to longer path lengths, the process also looks very different at the micro scale in the spatial and non-spatial models due to the localisation of the spread. In the spatial model, already at the beginning of the spread, the contacts of infectious nodes are more likely to occur with other infectious nodes than in the non-spatial model. This localisation of the spreading front hinders spreading: in a non-spatial model, contacts with infectious nodes other than a node's own infector are initially rare, and thus a larger fraction of contacts can transmit the spread. 

We found that the spatiotemporal framework is well-suited for modelling social distancing in a principled way, as a form of imposing spatial constraints on the temporal network structure. In particular, social distancing is an intervention measure aimed at reducing long-range weak links, and as such can be modelled in a straightforward manner by shrinking the interaction radii of nodes. We have observed that such spatially targeted interventions can be very effective even when the reduction in the total number of contacts is small, as opposed to reducing events by a similar amount in a non-spatial model. These results agree with empirical observations made during the COVID-19 pandemic (see, e.g.,~\cite{Schlosser20}) and underline both the importance and effectiveness of social distancing measures in epidemic control, as well as the importance of incorporating spatiality in models of interventions.

Interestingly, in our framework, space does not need to be directly interpreted as physical or geographical space, but may instead be considered as a latent embedding space of nodes. 
This further allows for high flexibility in model applications as well as generalisation to higher $n$-dimensional embedding spaces using $n$-spheres in place of 2D discs. The latent space allows us to capture the inherent social structure and homophily of individuals, or more generally, relations and similarities between data points. This point of view offers an additional layer of meaning to distance-based constraints: it is known that while social network structures are related to geospatial distances (e.g.,~\cite{Lambiotte2008,Onnela2011}), there is no one-to-one mapping between the social and geographic spaces.

Note that space implicitly produces memory in the system, as nodes in close proximity show a stronger tendency to be in repeated contact. This memory brings about social network traits such as link weight heterogeneity, heavier local triangles, and high clustering. There have been mechanisms developed to explicitly generate these traits and create memory in temporal networks, but here, these are induced by space.\\

\section{Appendix}

\textbf{Characteristic clustering value $c_{max}$}\\

Picking nodes $j,k$ uniformly randomly from the disc $D_R(i)$ centred at node $i$, $c_i$ is defined as the probability that $j,k$ are connected. That is, in the limit $T \rightarrow \infty$, the probability of their distance being at most $R$ can be written as 

\begin{equation}
    c_i \rightarrow P(d(j,k)\le R | j,k \in D_R(i)).
\end{equation}

Under the uniform spatial density assumption this can be expressed as the area of the lens formed by the intersection of interaction disks of $D_R(j),D_R(k)$ normalised by the area of $i$'s interaction disk:

\begin{equation}
    c_i \rightarrow \frac{Area(D_R(j) \cap D_R(k))}{\pi R^2}.
\end{equation}

The area of the lens $Area(D_R(j) \cap D_R(k)) := A(r)$ depends on the distance $d(j,k) := r$, and due to symmetry can be expressed as twice the area difference between a sector and an isosceles triangle as follows:

\begin{equation}
\begin{aligned}
     A(r)= 2 \left[ \frac{1}{2}R^2 \theta - \frac{1}{2}R^2 \sin\theta\right]
    \\=2R^2 \arccos\left( \frac{r}{2R} \right) - \frac{r}{2} \sqrt{4R^2 -r^2}
\end{aligned}
\end{equation}

where $\theta = 2 \arccos\left( \frac{r}{2R} \right)$. 

On the other hand, the pdf for $r$ under uniform density assumption within disc $D_R(i)$ is 
\begin{equation}
    f(r) = \frac{2r}{R^2} \text{ for } r \in [0,R].
\end{equation}

Therefore, the maximum attainable average local clustering coefficient can be computed as

\begin{equation}
    c_{max} = \int_0 ^R \frac{A(r) 2r}{\pi R^4} dr = 1- \frac{3 \sqrt{3}}{4 \pi} \approx 0.5865.
\end{equation}

In the limit $T \rightarrow \infty$ with the accumulation of links the spatiotemporal model converges to an RGG, and therefore the mean clustering coefficient converges to and attains this characteristic maximal value 

\begin{equation}
    c \rightarrow c_{max} =1- \frac{3 \sqrt{3}}{4 \pi} \approx 0.5865.
\end{equation}

This in fact agrees with the clustering result \cite{dall2002random} obtained for 2-dimensional random geometric graphs:

\begin{equation}
    c_2 = 1- H_2(1) = 1-\frac{\Gamma(1)}{\sqrt{\pi} \Gamma(1.5)} \left( \frac{3}{4} \right) ^{\frac{3}{2}} = 1- \frac{3 \sqrt{3}}{4 \pi}.
\end{equation}\\

\bibliography{citations}

\end{document}